\documentclass[usenatbib]{mn2e}

\usepackage{graphicx}

\usepackage{ifthen}

\def\ltsima{$\; \buildrel < \over \sim \;$}

\def\simlt{\lower.5ex\hbox{\ltsima}}

\def\gtsima{$\; \buildrel > \over \sim \;$}

\def\simgt{\lower.5ex\hbox{\gtsima}}

\def\kms{{\rm\,km\,s^{-1}}}

\def\kpc{{\rm\,kpc}}

\def\msun{{\rm\,M_\odot}}

\newcommand{\fmmm}[1]{\mbox{$#1$}}

\newcommand{\scnd}{\mbox{\fmmm{''}\hskip-0.3em .}}

\newcommand{\mcnd}{\mbox{\fmmm{'}\hskip-0.3em .}}

\def\deg{^\circ}

\def\s{\ifmmode \widetilde \else \~\fi}

\def\={\overline}

\def\spose#1{\hbox to 0pt{#1\hss}}

\def\lta{\mathrel{\spose{\lower 3pt\hbox{$\mathchar"218$}}

     \raise 2.0pt\hbox{$\mathchar"13C$}}}

\def\gta{\mathrel{\spose{\lower 3pt\hbox{$\mathchar"218$}}

     \raise 2.0pt\hbox{$\mathchar"13E$}}}

\def\Dt{\spose{\raise 1.5ex\hbox{\hskip3pt$\mathchar"201$}}}    

\def\dt{\spose{\raise 1.0ex\hbox{\hskip2pt$\mathchar"201$}}}    

\def\dotsfill{\leaders\hbox to 1em{\hss.\hss}\hfill}

\def\sun{\odot}

\def\Gyr{{\rm\,Gyr}}

\loadboldmathitalic

\title[Draco, a flawless dwarf galaxy]
{Draco, a flawless dwarf galaxy}

\author[Mathieu S\'egall, Rodrigo Ibata, Michael Irwin, N., F. Martin  \& Scott Chapman]
{M. S\'egall$^{1,*}$, R. A. Ibata$^{1}$, M. J. Irwin$^2$, N. F. Martin$^{1,3}$  \&
 S. Chapman$^2$ \\
$^{1}$ Observatoire de Strasbourg, 11, rue de l'Universit\'e, F-67000, Strasbourg, France\\
$^{2}$ Institute of Astronomy, Madingley Road, Cambridge, CB3 0HA,
  U.K. \\
$^{3}$ Max-Planck Institut f\"ur Astronomie, K\"onigstuhl 17, D--69117 Heidelberg, Germany}

\date{\today}

\begin{document} 

\maketitle 

\begin{abstract}
The Draco dwarf spheroidal galaxy (dSph), with its apparent immense mass to
light ratio and compact size, holds many clues to the nature of the
enigmatic dark matter.  Here we present deep photometric studies of this
dwarf galaxy, undertaken with the MegaCam Camera at the Canada-France-hawaii
Telescope, the Wide Field Camera at the Isaac Newton Telescope and the
Wide-Field and Planetary Camera on board the Hubble Space Telescope.  The
new photometric data cover the entirety of the galaxy, and reach $i'=24.5$
at 50\% completeness, significantly deeper than previous panoramic studies,
allowing searches for tidal disturbances of much lower surface brightness
than has been possible before.  With these improved statistics, we find no
evidence for asymmetric disturbances or tidal tails that possess more than
3\% of the stars found within the centre of the galaxy.  We find that the
central stellar density, as probed by the HST data, rises into the central
$0\mcnd5$. Uncertainties in the position of the centroid of the galaxy do
not allow us to determine whether the apparent flattening of the profile
interior to $0\mcnd5$ is reliable or not.  Draco is therefore a flawless
dwarf galaxy, featureless and apparently unaffected by Galactic tides.
\end{abstract}


\section{Introduction}

\renewcommand{\thefootnote}{\fnsymbol{footnote}} \footnotetext[1]{Based on
observations obtained with MegaPrime/MegaCam, a joint project of CFHT and
CEA/DAPNIA, at the Canada-France-Hawaii Telescope (CFHT) which is operated
by the National Research Council (NRC) of Canada, the Institute National des
Sciences de l'Univers of the Centre National de la Recherche Scientifique of
France, and the University of Hawaii. \\ Based on observations made with the
NASA/ESA Hubble Space Telescope, obtained from the data archive at the Space
Telescope Institute. STScI is operated by the association of Universities
for Research in Astronomy, Inc. under the NASA contract NAS 5-26555.}

The nature of the most common type of galaxy in the Local Group, the dwarf
spheroidals, remains a mystery.  Were these small galaxies formed early in
the Universe, as predicted in the hierarchical galaxy formation model
\citep{white}, and were many of them later incorporated into the Milky Way?
Though the $\Lambda$CDM model gives currently the best means to explain the
formation of these galaxies, this scenario has the well-known problem of a
discrepancy of 1--2 orders of magnitude between the number of low mass
structures predicted by this theory and the observed number of satellite
galaxies \citep{moore99a, klypin}. Their apparent alignments in great circles
\citep{lynden-Bell} have led some authors to propose a tidal
(non-cosmological) origin \citep{kroupa}, possibly by the early interaction
between Andromeda and the Milky Way \citep{sawa}.

Thus the study of dwarf spheroidal galaxies is of importance in our attempt
to understand the formation of galaxies in general and also the distribution
of the dark matter.  These small galaxies are most easily examined in the
Local Group due to their proximity, and of the dwarf spheroidal galaxies in
this region Draco is arguably the most interesting in terms of its
importance in placing constraints on the distribution of dark matter on
small scales.

The Draco dSph is a small, faint, and metal-poor Galactic satellite situated
about $71 \kpc$ from the Sun (see Table~1). The stellar population is very
old ($>$ 8-10 \Gyr) and very metal poor (${\rm [Fe/H]}=-1.8 \pm 0.2$~dex)
\citep{apparicio}.  Its total i-band luminosity is $(L/L_\sun) = 2.4 \pm 0.5
\times 10^5$ \citep{odenkirchen}, and King model \citep{king} fits to the
stellar profile, show that it has a core radius of 7.7\arcmin, and a tidal
radius of 40.1\arcmin.  Under the assumption of Virial equilibrium, the high
stellar velocity dispersion implies an extremely high mass to light ratio of
about $146 \pm 42$ in Solar units, according to \citet{odenkirchen}
(hereafter O01) or $330 \pm 125$ according to \citet{kleyna02} (hereafter
K02).  Several other studies (e.g., \citealt{klessen1}; \citealt{lokas02}),
also conclude that it is probably a very strongly dark-matter dominated
galaxy.

In many of its properties Draco is very similar to the Ursa Minor dSph,
another low-mass Milky Way satellite.  Surveys of the latter galaxy,
however, have shown the presence of tidally-induced features which attest to
the non-equilibrium state of that galaxy \citep{olszewski,bellazzini, palma03}, and
imply that the dark matter fraction may well be much lower in Ursa Minor
than might be naively expected. In contrast, previous studies of Draco have
shown it to have a smooth symmetrical shape (O01; \citealt{piatek,
bellazzini}).  Indeed, no study has yet found evidence for a distorted
distribution of stars in the halo of the Draco dSph, which would be expected
if it did not possess a massive dark matter mini-halo to protect the stellar
component from the strong tidal field of the Milky Way. Recently however,
\citet{wilkinson04} (hereafter W04) discovered a break in the light profile
of Draco at $\approx 25 \arcmin$ and a sharp decline in the velocity
dispersion at $\approx30 \arcmin$.  They concluded that the outer part of
Draco is filled with a dynamically colder stellar population, which could be
caused by an external tidal field.

Our aim in this work is therefore to search more thoroughly for tidal
perturbations in the halo of the Draco dSph, using the new generation of
wide-field cameras.  The outline of the paper is as follows: \S2 describes the
observations and the data reduction processes, \S3 details the selection
criteria for Draco stars, while \S4 discusses the analysis of the
photometric survey, the different star populations that are present, and the
search for substructures in the outskirts of this small galaxy.    
Finally, \S5 and \S6 are dedicated to the discussion of our results and
the conclusions of the study.

\begin{table}
\caption{Properties of Draco}
\label{tab1}
\begin{tabular}{lll}
\hline
\hline
 Parameter & Value & Reference \\
\hline
$\alpha$ (J2000) & $17^h20^m13.2^s$& O01 \\
$\delta$ (J2000) & $+57\deg 54' 54''$& O01 \\
(l,b) & ($86.4\deg$,$+34.7\deg$) & \\
d & $71\pm 7\kpc$ & O01 \\
$\rm{V_r}$ & $-293.8^{+2.6}_{-2.7}\kms$ & \citet{hargreaves} \\
$\sigma_{\rm{V}}$ & $10.5^{+2.2}_{-1.7}\kms$ & \citet{hargreaves} \\
$\rm{R_c}$ & $7.7'$ or $0.16\kpc$ & O01 \\
$\rm{R_t}$ & $40.1'$ or $0.83\kpc$ & O01 \\
E(B-V) & 0.029 & \\
$\rm{[Fe/H]}$ & $-1.8\pm 0.2$~dex & \citet{apparicio} \\
M & $3.5\pm 0.7 \times 10^7\msun$ & O01 \\
$\rm{(M/L)_i}$ & $146\pm 42$ & O01 \\
$\rm{M_V}$ & -8.8 & \citet{mateo} \\
\hline
\hline
\end{tabular}
\end{table}

\section{Observations and data reduction}

\subsection{MegaCam and INT surveys}

The Draco dSph galaxy was observed with the wide-field MegaCam camera at the
Canada-France-Hawaii Telescope (CFHT) in service mode during the 2004A and
2005A observing seasons.  Megacam is a mosaic of 36 ${\rm 4612\times 2048}$
EEV chips, offering a field of $\sim 0.96\times0.94$~square degrees. We
chose to observe seven fields, six around the galaxy and a comparison field
four degrees in declination below the centre of the galaxy.  Taken in
photometric conditions, the images totalled 950~s, 1090~s and 1700~s per
field in, respectively, the MegaCam g, r and i bands.  The seeing ranged
from $0\scnd46$ to $1\scnd12$, with the majority of the images being taken
in seeing better than $0\scnd8$.

The images were pre-processed by the CFHT "Elixir" pipeline yielding
debiassed, flat-fielded images, with i-band fringing artefacts removed.
Photometric standards observed over the season are used by "Elixir" to
measure the zero-point in each passband for every pointing. The resulting
images were subsequently passed through the photometry pipeline developed
for the Wide Field Camera (WFC) of the Isaac Newton Telescope (INT)
\citep{irwin2}, suitably altered to deal with MegaCam data.

Catalogue generation follows the precepts outlined by \citet{irwin85,
irwin96} and includes the facility to: automatically track any background
variations on scales of typically ~20-30 arcsec; detect and deblend images
or groups of images; and parameterise the detected images to give various
(soft-edged) aperture fluxes, position and shape measures.  The generated
catalogues start with an approximate World Coordinate System (WCS) defined
by the known telescope and camera properties (the WCS distortion model) and
are then progressively refined using the 2MASS (2 Micron All Sky Survey) astrometric
catalogues to give an internal precision generally better than 0.1 arcsec
and a global external precision of ~0.25 arcsec.  These latter numbers are
solely dependent on the accuracy of the astrometric catalogues used in the
refinement.

All catalogues for all CCDs for each pointing are then processed using the
image shape parameters for morphological classification in the main
categories: stellar, non-stellar, and noise-like.  A sampled curve-of-growth
for each detected object is derived from a series of aperture flux measures
as a function of radius.  The classification is then based on comparing the
curve-of-growth of the flux for each detected object with the well-defined
curve-of-growth for the general stellar locus.  This latter is a direct
measure of the integral of the point spread function (PSF) out to various
radii and is independent of magnitude, if the data are properly linearised,
and if saturated images are excluded.  The average stellar locus on each
detector is clearly defined and is used as the basis for a null hypothesis
stellar test for use in classification.  The curve-of-growth for stellar
images is also used to automatically estimate frame-based aperture
corrections for conversion to total flux.

Various quality control plots are generated by the pipeline and these were
used to monitor characteristics such as the seeing, the average stellar
image ellipticity (to measure trailing), the sky brightness and sky noise,
the size of aperture correction for use with the ``optimal'' aperture flux
estimates (here "optimal" refers to the well-known property that soft-edged
apertures of roughly the average seeing radius provide close to profile fit
accuracy, e.g., \citealt{naylor98}).  The "optimal" catalogue fluxes for the
g, r, i filters for each field were then combined to produce a single matched
catalogue for each pointing and the overlaps between pointings were used to
cross-calibrate all the catalogues to a common system to within an accuracy
$\sim 1$\% across the survey region.

For completeness we also present here an earlier panoramic dataset obtained
with the INT WFC on the nights of the 11th, and the 13th to 15th of October
2002.  The WFC is a mosaic of four ${\rm 4k\times 2k}$ EEV chips, offering a
field of $\sim 0.29$~square degrees.  A total of 11 pointings were observed
in photometric conditions, with exposures of 1800~sec in the equivalent of
the Johnson V passband and 1800~sec in the equivalent of the Gunn i band at
each pointing.  The seeing ranged from $0\scnd7$ to $1\scnd1$.  The INT WFC
images were reduced in an identical manner to the MegaCam observations, with
the exception that the pre-processing was performed with the INT pipeline.
A small offset of $0.106$ magnitudes was applied to the i-band INT magnitudes
to bring them into agreement with the MegaCam values.

Tables~\ref{tab2} and~\ref{tab3} list the photometry resulting from the
MegaCam and INT surveys, showing g, r, i-band MegaCam magnitudes
(in the AB system), V and
i-band INT magnitudes (converted into the MegaCam AB system),
with their uncertainties, a classification parameter
in each passband ($c_g$, $c_r$, $c_i$ and $c_V$), 
and the equatorial coordinates for each object. This
classification index is a flag that indicates the likely nature of the
objects. A classification index of 1 corresponds to a likely galaxy,
$-1,\dots, -9$ to a point-sources (in order of decreasing confidence of
stellarity), and 0 to an artefact or noise.  The INT photometry is only
used to calibrate the HST V-band photometry.

\begin{table*}
\caption{The first 10 rows of the MegaCam catalogue. Magnitude values of $0.000$
indicate that no measurement was possible.}
\label{tab2}
\begin{tabular}{ccccccccccc}
\hline
\hline
$\alpha$ & $\delta$ & $\rm{g}$ & $\delta\rm{g}$ & $\rm{c_g}$ &
$\rm{r}$ & $\delta\rm{r}$ & $\rm{c_r}$ & $\rm{i}$ &
$\delta\rm{i}$ & $\rm{c_i}$ \\
$(^{h~m~s})$ & $(\deg~'~'')$ & & & & & & & & & \\
\hline
 17 16 45.50 & 53 24 41.1 &  0.000 & 0.000 &  0 &  0.000 & 0.000 &  0 & 23.406 & 0.057 &  1 \\                                      
 17 16 44.52 & 53 24 41.1 & 24.204 & 0.051 &  1 & 23.708 & 0.054 &  1 &  0.000 & 0.000 &  0 \\                                      
 17 17  3.98 & 53 24 41.4 & 24.394 & 0.060 &  0 & 22.796 & 0.025 &  0 & 21.893 & 0.016 &  0 \\                                      
 17 17  1.92 & 53 24 41.7 & 19.279 & 0.002 &  1 & 18.065 & 0.001 & -1 & 17.304 & 0.001 & -1 \\                                      
 17 17 17.34 & 53 24 42.4 & 24.223 & 0.052 &  1 & 23.060 & 0.031 & -8 & 22.391 & 0.024 & -8 \\                                      
 17 18 10.29 & 53 24 42.4 &  0.000 & 0.000 &  0 &  0.000 & 0.000 &  0 & 22.387 & 0.024 &  1 \\                                      
 17 18 10.58 & 53 24 42.4 &  0.000 & 0.000 &  0 &  0.000 & 0.000 &  0 & 22.310 & 0.023 &  0 \\                                      
 17 17 33.88 & 53 24 42.5 &  0.000 & 0.000 &  0 &  0.000 & 0.000 &  0 & 34.916 & 8.326 &  0 \\                                      
 17 18 13.85 & 53 24 42.5 &  0.000 & 0.000 &  0 &  0.000 & 0.000 &  0 & 22.317 & 0.023 &  0 \\                                      
 17 18 14.27 & 53 24 42.5 &  0.000 & 0.000 &  0 &  0.000 & 0.000 &  0 & 22.334 & 0.023 &  1 \\                                      
\hline
\hline
\end{tabular}
\end{table*}

\begin{table*}
\caption{The first 10 rows of the INT catalogue.}
\label{tab3}
\begin{tabular}{ccrrrrrr}
\hline
\hline
$\alpha$ & $\delta$ & $\rm{V}$ & $\delta\rm{V}$ & $\rm{c_V}$ &
$\rm{i}$ & $\delta\rm{i}$ & $\rm{c_i}$ \\
$(^{h~m~s})$ & $(\deg~'~'')$ & & & & \\
\hline
17 20 17.72 & 57  4 45.4 & 22.572 &  0.034 &  1 & 21.026 &  0.020 &  1 \\
17 20 16.05 & 57  4 47.0 & 23.862 &  0.105 &  0 &  0.000 &  0.000 &  0 \\
17 19 53.57 & 57  4 47.4 &  0.000 &  0.000 &  0 & 23.099 &  0.121 &  0 \\
17 20 18.82 & 57  4 47.6 &  0.000 &  0.000 &  0 & 22.936 &  0.105 & -3 \\
17 19 55.91 & 57  4 48.0 &  0.000 &  0.000 &  0 & 22.627 &  0.080 &  1 \\
17 20  5.51 & 57  4 48.6 & 23.617 &  0.085 &  0 &  0.000 &  0.000 &  0 \\
17 19 47.26 & 57  4 48.7 & 21.466 &  0.013 & -3 & 19.815 &  0.007 & -3 \\
17 19 26.95 & 57  4 48.8 & 23.799 &  0.100 & -2 &  0.000 &  0.000 &  0 \\
17 19 19.21 & 57  4 49.5 &  0.000 &  0.000 &  0 & 23.262 &  0.140 &  1 \\
17 19 11.11 & 57  4 50.0 & 22.231 &  0.025 & -3 & 21.859 &  0.041 & -3 \\
\hline
\hline
\end{tabular}
\end{table*}

\subsection{HST survey}

We also analysed a series of HST images of a field in the centre of the
Draco dSph.  The data were taken in 1995 and 1999 with the Wide Field and
Planetary Camera (WFPC2), in two spectral bands, V (F606W) and i (F814W).
The total exposure time amounts to 4000~sec in F606W and 4500~sec in F814W.
The wide-field (WF) chips of the mosaic were reduced with the
DAOPHOT/ALLSTAR package \citep{stetson87,stetson90}.  A point spread function
(PSF) previously constructed for a study of the globular cluster M4
\citep{ibata99} in each of the WF chips was used to fit the profiles of all
detected sources in the WF mosaic.  This yielded, for each detected object,
the coordinates, the magnitude, a parameter describing the profile shape
(sharpness) and the $\chi$ value of the PSF fit. The equatorial coordinates
of the sources were determined by comparison to the MegaCam survey.  The
instrumental WFPC2 F606W and F814W magnitudes were calibrated onto the INT
V-band and MegaCam i-band systems, respectively, by assuming a simple offset
with no colour terms. The correspondence between the bands is reasonably
acceptable, as we show in Fig.~1.

\begin{figure}
\begin{center}
\includegraphics[width=\hsize]{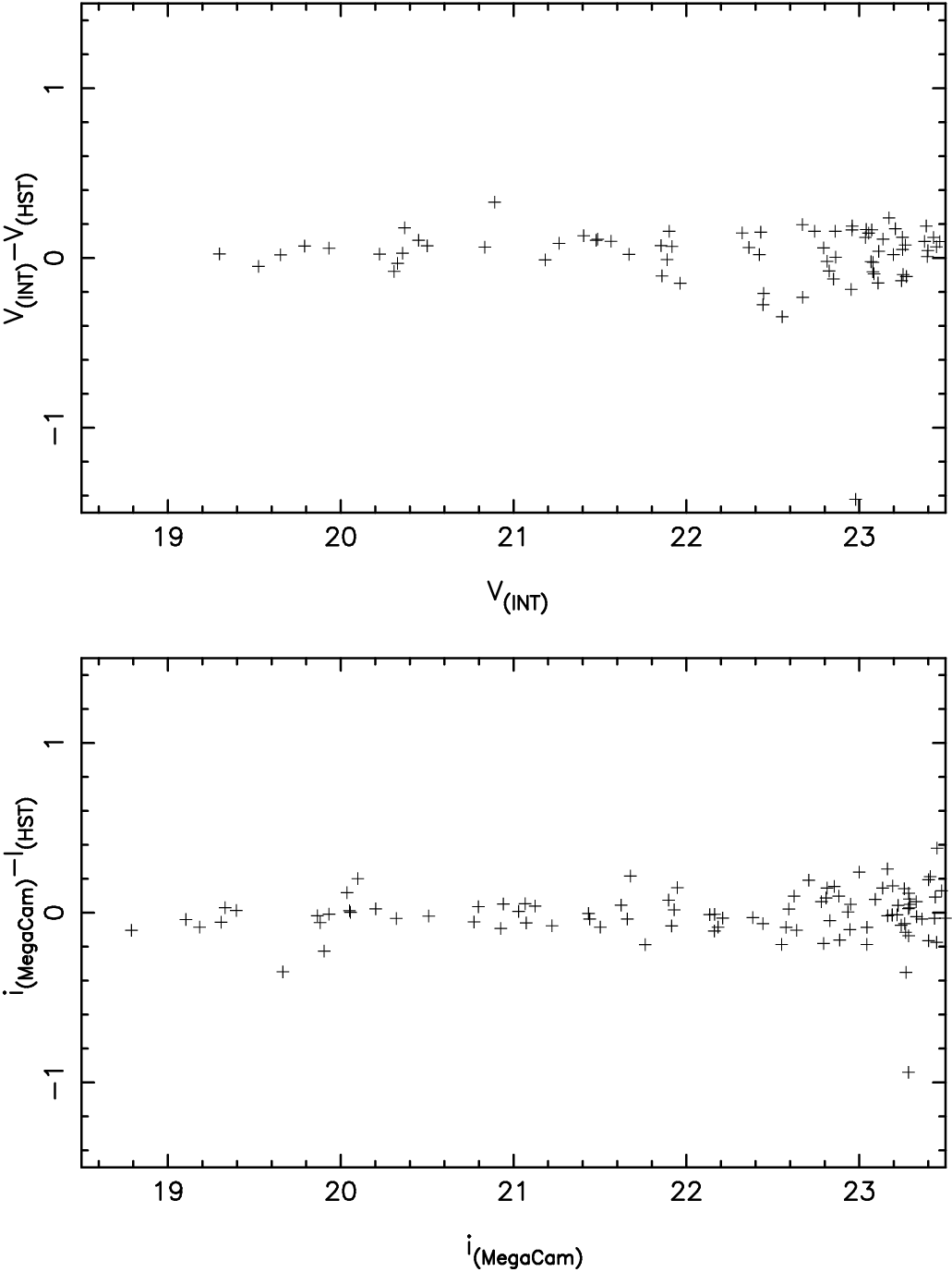}
\caption{HST F606W photometry was calibrated as a simple offset
from the INT V band (upper panel). Likewise, HST F814W photometry 
was calibrated with MegaCam i (lower panel). 
(The increased scatter at i=20 is due to photometric variability
of horizontal branch stars).}
\end{center}
\end{figure}

The images listed in Table~5 do not cover exactly the same area of sky, and
have offsets of up to $\sim 80$~pixels.  Since we were primarily interested
in measuring the stellar density profile in the inner regions of the Draco
dSph, we retained only the overlapping region in common to all 14 exposures,
thereby ensuring that the region was surveyed to a uniform depth.

\section{Sample selection}

\subsection{Completeness}

The deep HST photometry in the centre of the galaxy is useful to help define
the stellarity criteria and also to determine the completeness of the
MegaCam survey. 
We find that adopting MegaCam sources with classification indices of either
-1 or -2 in all three passbands results in zero false objects (i.e. zero
sources not present in the HST survey).  The completness for this selection
is represented with a dashed line in Fig.~2. We also show, with a
solid line, the completness when we select all the MegaCam objects
(classification index from 1 to -9). To $i=24$ both selections have
completeness $> 50$\%. Down to this limit, the adopted selection on
classification indices does not make us lose a substantial fraction of
stars. It should be noted that the incompleteness is affected to a
significant extent by the crowding, which is maximal in the central region
probed by the HST field, but will be much less problematic in the outskirts
of the galaxy. Fig.~2 therefore represents the worst completeness over the survey.

\begin{figure}
\begin{center}
\includegraphics[angle=-90,width=\hsize]{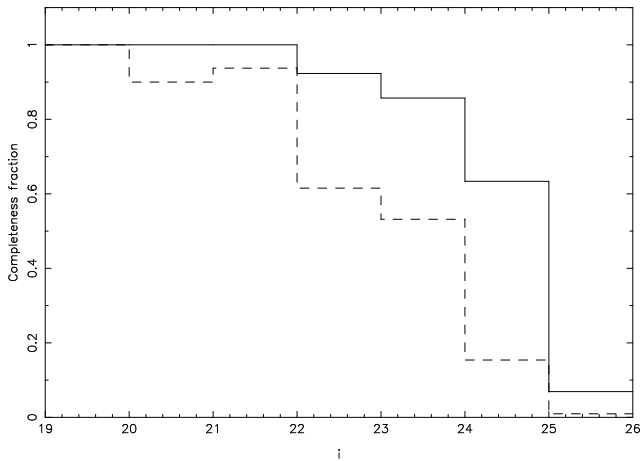}
\caption{The completeness of the MegaCam dataset is
compared to the HST survey for all the catalogue (solid line) and
for the classification indices -1 and -2 (dashed line).}
\label{fig02}
\end{center}
\end{figure}

\begin{table}
\caption{HST field IDs}
\label{tab4}
\begin{tabular}{llrr}
\hline
\hline
name & passband & exposed time & date \\
 & & (s) & \\
\hline
u2oc0101t & F606W & 200 & 1995 \\
u2oc0102t & F606W & 1100 & 1995 \\
u2oc0103t & F606W & 1000 & 1995 \\
u2oc0104t & F814W & 200 & 1995 \\
u2oc0105t & F814W & 1100 & 1995 \\
u2oc0106t & F814W & 1100 & 1995 \\
u5er1401r & F606W & 400 & 1999 \\
u5er1402r & F606W & 400 & 1999 \\
u5er1403r & F606W & 500 & 1999 \\
u5er1404r & F606W & 500 & 1999 \\
u5er1405r & F814W & 400 & 1999 \\
u5er1406r & F814W & 500 & 1999 \\
u5er1407r & F814W & 500 & 1999 \\
u5er1408r & F814W & 500 & 1999 \\
\hline
\hline
\end{tabular}
\end{table}

\begin{table*}
\caption{The first 10 rows of the HST catalogue.}
\label{tab5}
\begin{tabular}{ccrrrrrrrr}
\hline
\hline
$\alpha$ & $\delta$ & $\rm{V}$ & $\delta\rm{V}$ & $\rm{V_{\chi}}$ &
$\rm{V_{sharp}}$ & $\rm{i}$ & $\delta\rm{i}$ & $\rm{i_{\chi}}$ &
$\rm{i_{sharp}}$\\ 
(J2000) & (J2000) & & & & & & & & \\
\hline
17 20  3.41 &  57 53 31.3 & 26.538 &  0.146 &  0.850 &  0.048 & 26.158 &  0.185 &  0.670 &  0.040 \\
17 20  3.28 &  57 53 32.8 & 24.845 &  0.031 &  1.060 &  0.010 & 24.616 &  0.049 &  1.270 &  0.162 \\
17 20  3.31 &  57 53 33.4 & 24.048 &  0.017 &  1.920 & -0.141 & 23.615 &  0.022 &  3.450 &  0.041 \\
17 20  3.14 &  57 53 33.5 & 23.702 &  0.014 &  0.560 &  0.081 & 23.602 &  0.023 &  1.060 &  0.109 \\
17 20  3.51 &  57 53 34.8 & 25.259 &  0.045 &  0.510 &  0.118 & 24.986 &  0.067 &  0.800 &  0.080 \\
17 20  3.02 &  57 53 34.9 & 25.801 &  0.064 &  0.660 &  0.081 & 25.464 &  0.089 &  0.630 &  0.079 \\
17 20  3.71 &  57 53 34.9 & 25.907 &  0.068 &  0.650 &  0.053 & 26.065 &  0.143 &  0.650 & -0.014 \\
17 20  2.71 &  57 53 36.2 & 27.108 &  0.206 &  0.590 & -0.048 & 26.180 &  0.165 &  0.930 &  0.056 \\
17 20  3.98 &  57 53 36.7 & 25.604 &  0.062 &  1.330 &  0.099 & 25.304 &  0.075 &  1.600 &  0.177 \\
17 20  2.82 &  57 53 36.9 & 24.948 &  0.031 &  0.760 &  0.062 & 24.806 &  0.050 &  0.950 &  0.136 \\
\hline
\hline
\end{tabular}
\end{table*}

\subsection{Star selection}

\begin{figure}
\begin{center}
\includegraphics[angle=-90,width=\hsize]{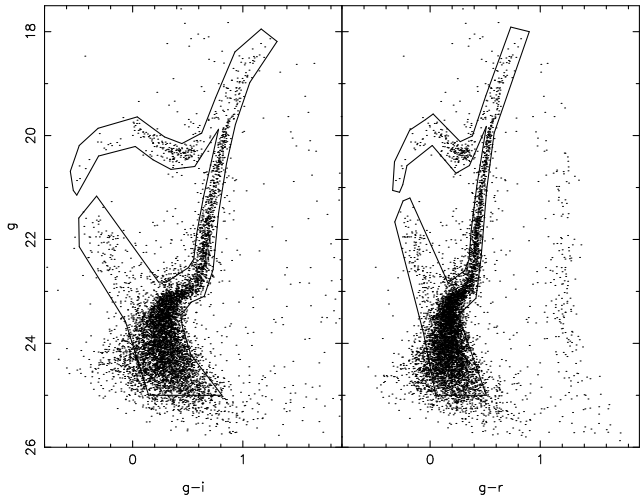}
\caption{The colour-magnitude diagram of the central 6$\arcmin$ of the
Draco dSph, showing g magnitude as a function of $g-i$ (left) and $g-r$ (right).
The initial CMD selection is indicated.}
\label{fig03}
\end{center}
\end{figure}

Fig.~\ref{fig03} shows the colour magnitude diagram (CMD) of the MegaCam
stars in the central region of Draco within a radius of 6$\arcmin$. To
reduce the contamination by foreground stars and background galaxies, we
construct a selection region covering the red giant branch (RGB) to
sub-giant branch (SGB) on the central CMD. As the red giant branch has very
low contrast compared to the "background" in some fields (or may simply not
be present), this CMD selection will help reduce contaminants in the
analysis below. A fairly broad selection is chosen here to ensure that we do not
reject a significant fraction of members of the dwarf galaxy.

\begin{figure}
\begin{center}
\includegraphics[angle=-90,width=\hsize]{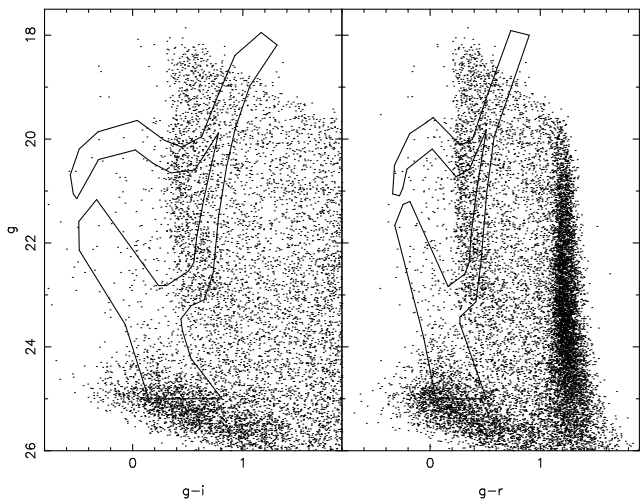}
\caption{As Fig.~3, but for the comparison field.}
\label{fig04}
\end{center}
\end{figure}

Clearly a simple CMD filter is not an optimal way to select Draco-like
stars, particularly towards fainter magnitudes, where the discrimination
between stars of the dwarf galaxy and background galaxies is difficult.  We
therefore try to implement a CMD filter that will enhance the signal of the dwarf 
galaxy by filtering out those CMD regions that degrade
the total signal to noise. To this end we take the 
background field (Fig.~4) as representative of the contamination by foreground 
and background Galactic stars as well as point-like distant galaxies that
have managed to evade the stellarity criterion.
In general, the signal to noise ratio in a given CMD bin will be simply
$$
S/N={{n_R - f n_{B}} \over {\sqrt{ \delta n_R^2 + f^2 \delta n_{B}^2 }}} ,
$$
where $n_R$ is the number of stars in the CMD bin in the region under consideration, $n_B$
is the number of stars in the comparison field, and $f$ is the ratio of the geometrical areas of the 
two fields. Assuming Poisson noise, this simplifies further to
$$
S/N={{n_R - f n_{B}} \over {\sqrt{ n_R + f^2 n_B }}} .
$$
Let us consider first the region inside $6\arcmin$ (displayed in Fig.~3),
taking the stars within the selection polygon.
We bin the CMD in intervals of $0.05$~mag
in g-i colour and $0.25$~mag in g-band magnitude. For each bin, we calculate the
signal-to-noise ratio, and we sort the resulting array. Starting with the CMD bin of
highest signal to noise, we examine each new bin in decreasing order of S/N testing whether
the additional bin adds to the total S/N, in which case the bin is accepted. 
It transpires that upon applying this process we accept the whole region within the 
polygon of Fig.~3, i.e. in the region within $6\arcmin$ of the galaxy centre, all of the CMD bins 
contribute usefully to the total S/N.

However, this will not the case on the outskirts of the galaxy, where $n_R - f n_B \approx 0$.
In the outer regions the signal from the dwarf galaxy is of course extremely weak, 
so one cannot apply the same 
procedure as above. We use instead the CMD within $6\arcmin$ as a model. 
Now the signal to noise ratio becomes:
$$
S/N={{g n_D } \over {\sqrt{  n_R + g^2 n_D }}} 
$$
where $n_D$ is the number of Draco stars in the central ($6\arcmin$) region and $g$ 
is an unknown normalisation factor combining the ratio
of geometrical areas between the area of interest and the central region and the relative
frequency of stars from the dwarf galaxy in the outer region compared to the central region.
Since $n_R >> g^2 n_D$ in this background-dominated case, we find that
$S/N \approx C  n_D  (n_R)^{-1/2}$, with $C$ a constant of proportionality whose value we need
not know. We proceed as above, binning the CMD, calculating the signal-to-noise ratio
in each bin (to within the multiplicative constant), and sorting the resulting array. 
We find that in the outskirts of the galaxy at radii beyond $45\arcmin$ 
an optimal CMD selection includes the main-sequence
turnoff and parts of the horizontal branch and blue-straggler regions.

The left and middle panels of Fig~5 show the result of these two selection methods,
where we present the full dataset of point-sources, filtered by the two CMD filters. 
These two CMD filters allow us to construct a star selection appropriate for
the centre of the dwarf galaxy (S1, on the left-hand panel), and the other 
(S2, in the middle panel) that enhances the
signal to noise in the external parts of the galaxy, where the background is
high.  We retain 25670 stars for the S1 selection and 5385 stars for S2.
Fig.~6 shows the spatial distribution of the two selections.
 
The matched-filter technique (see e.g., \citealt{rockosi}) provides an alternative 
means to enhance the signal of an expected CMD structure 
in the presence of contamination. We apply this procedure also
to our survey, taking the stars within the polygon of Fig.~3 to be representative of the 
desired signal, with a background sampled by the comparison field shown in Fig.~4.
The matched-filter method assigns a relative weight to each star, according to its 
CMD location (the map of the matched-filter weights is displayed in the 
right-hand panel of Fig.~5). In this way it is more efficient than the methods used 
above to select the samples "S1" and "S2" since one can make use of the entire 
dataset. However, the drawback is that one must bin all quantities of interest, which is not always convenient. We therefore use all three methods, applying each where appropriate.

Fig.~\ref{fig07} shows the selection boxes used to investigate the spatial
distribution of stars at other evolutionary phases which are clearly present
from visual inspection of the CMD: the blue and red horizontal branch
(hereafter BHB and RHB respectively), and the blue stragglers (hereafter
BS). These boxes are defined from the (g-i,g) CMD.  The spatial distribution
of these selections are shown on Fig.~\ref{fig08}, where we display maps for
the BHB, RHB and BS populations, which contain 605, 2542 and 973 stars,
respectively.

\begin{figure}
\begin{center}
\includegraphics[angle=-90,width=\hsize]{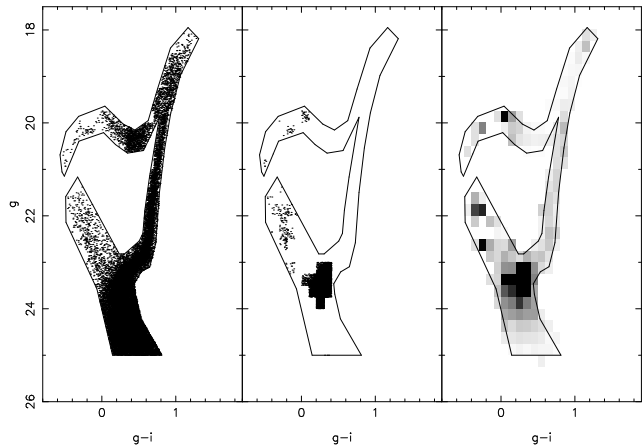}
\caption{The left and middle panels show, respectively, the result of the
CMD filters constructed to optimise signal to noise in the central regions
(S1) and in the outer regions of the galaxy (S2).  The right hand panel
shows the weight array of the matched filter analysis.}
\label{fig05}
\end{center}
\end{figure}

\begin{figure}
\begin{center}
\includegraphics[angle=-90,width=\hsize]{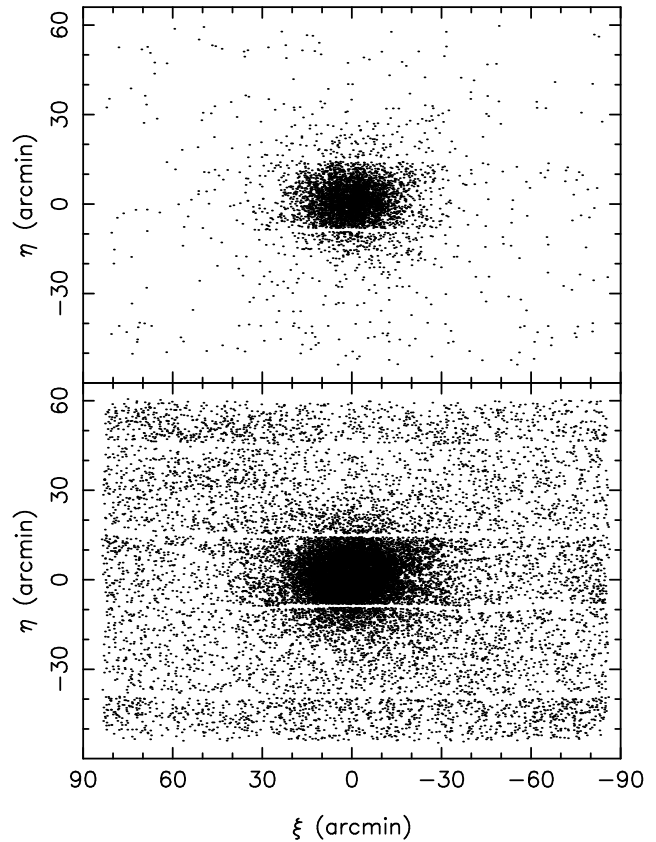}
\caption{The distribution of sources selected from the optimized CMD
selection regions displayed in Fig.~\ref{fig05}; the external optimisation
S2 is shown on the upper panel, whereas the central optimisation S1 is
portrayed on the lower panel. The positions of the sources are shown in
standard coordinates (gnomonic projection).}
\label{fig06}
\end{center}
\end{figure}

\begin{figure}
\begin{center}
\includegraphics[angle=-90,width=\hsize]{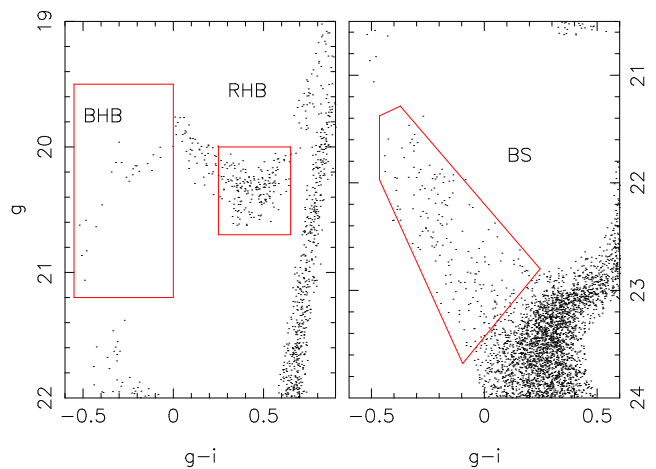}
\caption{The selection boxes for the BHB, RHB and BS evolutionary phases,
superimposed on the photometry of stars drawn from within the central $6\arcmin$.}
\label{fig07}
\end{center}
\end{figure}

\begin{figure}
\begin{center}
\includegraphics[angle=-90,width=\hsize]{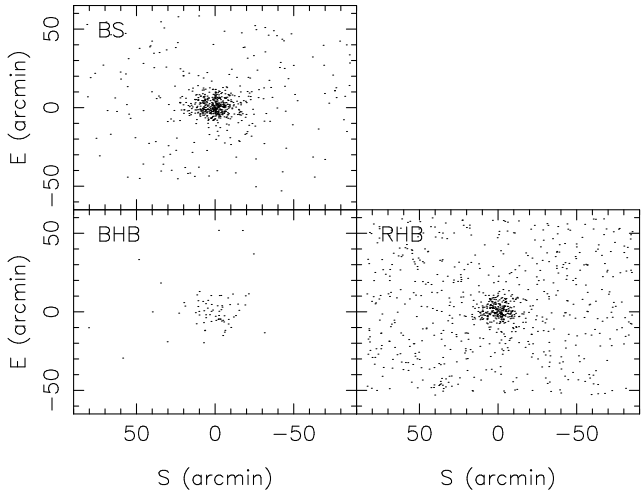}
\caption{The spatial distributions of the three selections, BHB, RHB and
BS, are shown.}
\label{fig08}
\end{center}
\end{figure}

\section{Analysis}

\subsection{Stellar density profile}

We next construct the stellar density profiles for the two
selections. However, there are two practical problems that need to be
solved.  First, the survey is riddled with holes caused by CCD artefacts
around bright stars, so these unusable areas of the fields need to be masked
out.  We eliminated circular regions around the positions of bright USNO and
Hipparcos stars which have magnitudes ${\rm B} < 16$, choosing appropriate
magnitude-dependent radii to cover the holes in the MegaCam survey. The
second problem is the gaps between CCDs. The chipsets of the MegaCam camera
are distributed on four rows, with two large gaps. These gaps (which are clearly visible in Fig.~6)
are sufficiently large to influence measurements of the stellar density profile.

To trace the stellar density profiles, we construct a succession of ellipses
with different major axis values (throughout this work, we use $s$ to denote
an elliptical radius, and $r$ to denote a circular radius).  We choose the
best shape parameters found by O01, a ellipticity equal to $0.30$ and a position
angle of the major axis of $88\deg$. For each optimised selection, we
construct the surface brightness profile, taking into account the holes made
by the bright stars and the gaps between CCDs, and remove a background
level. The background count level is determined from the comparison
field. In Fig.~9 we show the resulting profiles before background subtraction (upper panel)
and after background subtraction (lower panel).

\begin{figure}
\begin{center}
\includegraphics[width=\hsize]{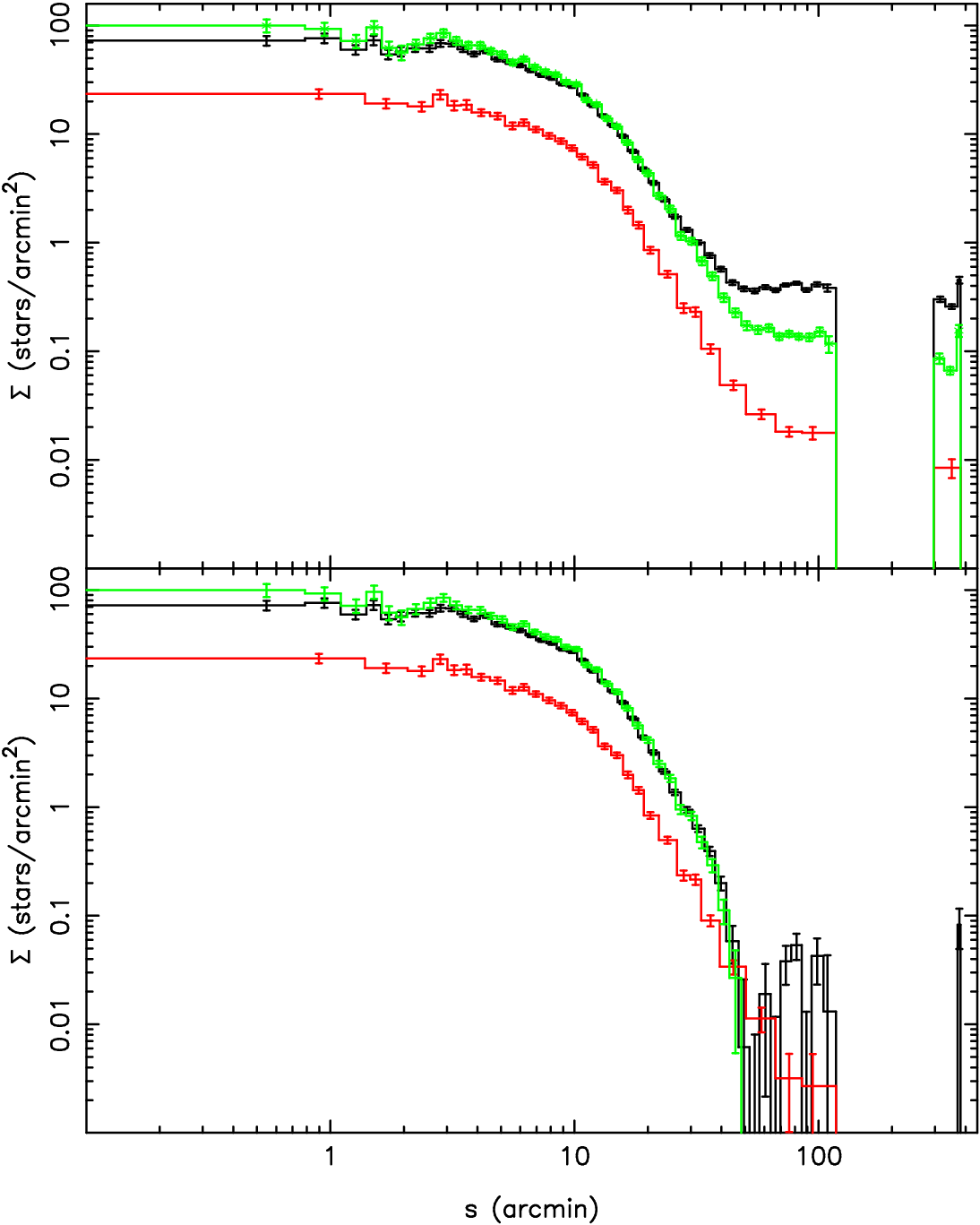}
\caption{The stellar profile of the Draco dSph galaxy. The upper panel shows
the profiles prior to background subtraction, while in the lower panel the
mean count level from the comparison field has been removed. The black histogram shows the profile of
the S1 sample, the S2 sample is displayed in red, and the profile resulting
from the matched-filter analysis is shown in green.}
\label{fig09}
\end{center}
\end{figure}

We  next fit  structural  models to  the  S1 and S2 profiles  and compare  our
results with O01 and \citealt{wilkinson02}  (hereafter W02). We
use three different models, the King model \citep{king}:
\[
\Sigma(s)=\Sigma_{0} \Bigg(\frac{1}{\sqrt{1+(s/R_{c})^{2}}}-\frac{1}{\sqrt{1+(R_{t}/R_{c})^{2}}} \Bigg)^{2}
\]
the exponential model \citep{sersic}:
\[
\Sigma(s)=\Sigma_{0} \exp(-(s/R_{0})^{n})
\]
and the Plummer model:
\[
\Sigma(s)=\Sigma_{0} R_{0}/(1+(s/R_{0})^{2})^{2} \, .
\]

\begin{figure}
\begin{center}
\includegraphics[width=\hsize]{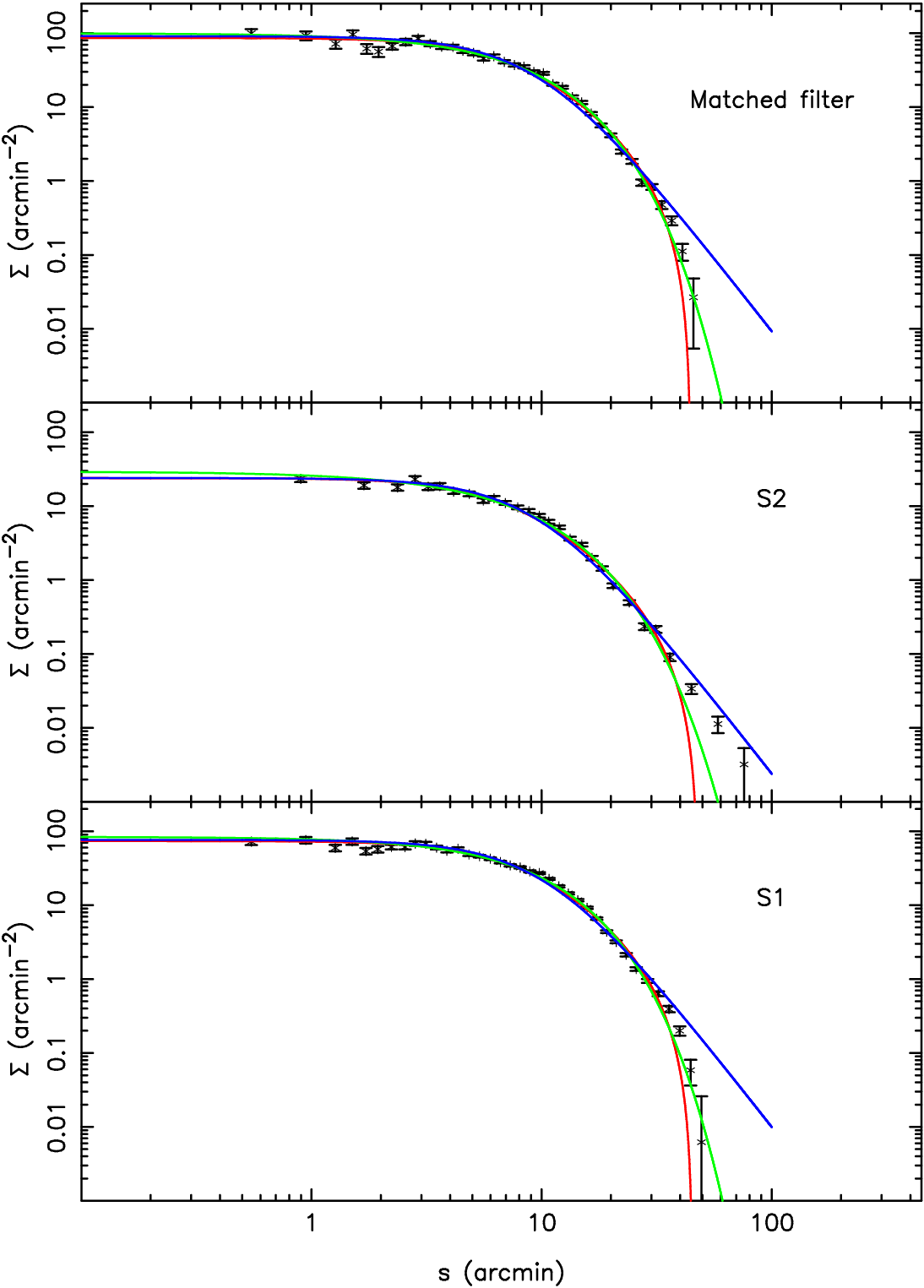}
\caption{Fits of a King (red), an exponential (green), and a Plummer (blue)
model to the stellar density profiles derived from the Matched-Filter sample (upper panel)
the S2 sample (middle panel) and the S1 sample (lower panel). The error  bars represent 1$\sigma$ uncertainties.}
\label{fig10}
\end{center}
\end{figure}

The fitting is performed by $\chi^2$ minimisation, with the ellipticity and
the position angle of the major axis taken as constant.  The results are
summarised in table~6, and the corresponding profiles are displayed in
Fig.~\ref{fig10}.

\begin{table}
\caption{Model fits for the parameters of the  King, exponential, and
Plummer profiles using the S1, S2 and Matched-Filter (MF) samples.
The quantity $\chi^2/(N-f)$ gives the reduced-$\chi^2$ for the fit.
We also indicate the values found in earlier studies.}
\label{tab6}
\begin{tabular}{ccccc}
\hline
\hline
 King & data & r$_{c} (\arcmin)$ & r$_{t}(\arcmin)$ & $\chi^2/(N-f)$ \\
\hline
 & S1 & $8.15 \pm 0.02$ & $45.0\pm 0.4$  & 5.35 \\
 & S2 & $7.18\pm 0.05$ & $48.9\pm 0.9$ & 4.61\\
 & MF & $7.63\pm 0.04$ & $45.1\pm 0.6$ & 3.31\\
 & O01 & $7.7\pm 0.2$ & $40.1\pm 0.9$ & \\
\hline
\hline
Exponential & data & r$_{0}(\arcmin)$ & n & $\chi^2$ \\
\hline
 & S1 & $8.07\pm 0.05$ & $1.20\pm 0.01$ & 3.97 \\
 & S2 & $6.80\pm 0.11$ & $1.07\pm 0.01$ & 3.42 \\
 & MF & $7.43\pm 0.10$ & $1.15\pm 0.01$ & 2.27 \\
 & O01 & $7.6\pm 0.1$ & $1.2\pm 0.1$ & \\
\hline
\hline
Plummer & data & r$_{0}(\arcmin)$  & & $\chi^2$ \\
\hline
 & S1 & $10.73\pm 0.01$ &  & 9.55 \\
 & S2 & $10.09\pm 0.02$ &  & 5.03 \\
 & MF & $10.06\pm 0.01$ &  & 8.01 \\
 & W02 & $9.71\pm 0.1$ &  & \\
\hline
\hline
\end{tabular}
\end{table}

The extensive spatial coverage of our survey provides us with information
far from the core of Draco, up to $\approx$ 100\arcmin. In contrast to W04,
who found a break at $25\arcmin$ in the light profile with the INT dataset,
the profiles on Fig.~\ref{fig09} based on either
the S2 selection or from the matched-filter analysis
do not show evidence for a break.  
 We suspect that the presence of the break in the
W04 study could have been an artefact of background subtraction, which is
very difficult at these low count levels.

The fits of Fig.~10 give a visual impression that the data can be 
represented quite well by these simple models. However, the reduced $\chi^2$
values listed in Table~6, are very poor, and even the best of the models
(the exponential model fit to the Matched-Filter sample) can be rejected
with better than 99.99\% confidence. Thus the data are now of such excellent
quality that these simple models no longer provide satisfactory fits.
We suspect that dynamically self-consistent triaxial models will be required
to reproduce the structure of this dwarf galaxy,
however that is beyond the scope of the present article.

\begin{figure}
\begin{center}
\includegraphics[angle=270,width=\hsize]{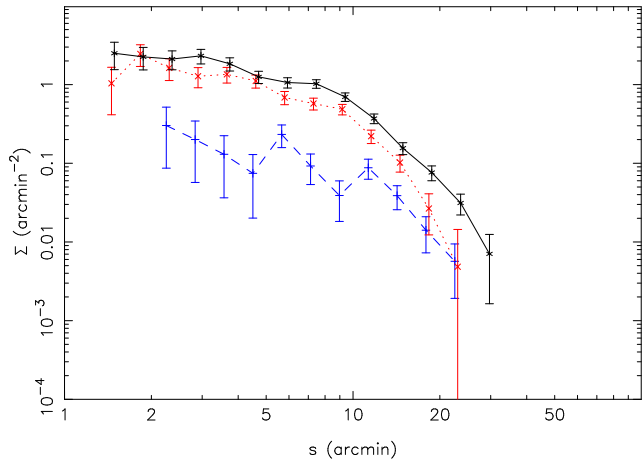}
\caption{The density profiles of the BHB (blue dashed line), RHB (red dotted line) and BS (black solid line) selections.}
\label{fig11}
\end{center}
\end{figure}

We also trace the stellar density profile for the BHB, RHB and BS
selections.  Fig.~\ref{fig11} shows the three resulting profiles. A fit with
a Plummer model yields scale radii of 
$11\mcnd8 \pm 3\mcnd8$, 
$7\mcnd6 \pm 0\mcnd3$ and
$8\mcnd1 \pm 0\mcnd2$ 
respectively for the BHB, RHB and BS selections. 
We find that widening the BHB box by allowing redder horizontal-branch stars into the selection
results in a profile that is closer to that of the RHB. Thus the paucity of BHB stars does not allow us to confirm the visual impression
from Fig.~8 that the BHB stars are more dispersed than the RHB and BS stars.
Consequently, Draco, unlike
the Sculptor \citep{tolstoy} and Fornax \citep{battaglia} dSph, does not possess a
BHB population that is significantly more extended 
than the RHB population. Whether this attests to a difference in early star-formation or
simply to the low statistics of BHB stars is currently not clear.

\subsection{Smoothness of the stellar density profile}

The stellar profiles displayed in Fig.~10 appear approximately smooth, but are they really featureless apart from their monotonic decrease? To investigate this point
we implement a variant of a test devised by \citet{kuhn} (hereafter KKF), who in a preliminary study
of the stellar profile of the Draco dSph, found the very interesting result that the inner regions within 
$20\arcmin$ are not as smooth as one would expect from Poisson noise statistics.
Following KKF, we divide our Matched-Filter sample into 1000 radial bins, using the same elliptical coordinate as in Fig.~10. Each bin contains 24 stars to maintain uniform
statistical significance. However, since our survey suffers from gaps between CCDs and holes due to the haloes of bright stars, we calculate the density $\rho_i$ in each bin $i$, and also the density uncertainty $\delta\rho_i$, taking into account the Poisson noise in each bin, and the Poisson noise in the background subtraction. 

In a similar way to KKF, we take groups of 5 bins at radius $R_i$, and calculate the 
reduced $\chi^2$ of the group, testing the null-hypothesis that the density is constant over the small radial interval. To this end, we define:
$$\chi_5^2 (R_i) = {{1}\over{5}} \sum_{j=0}^4 {{(\rho_i - \overline{\rho})^2}\over{\delta\rho_i^2}}$$
where $\overline{\rho}$ is the weighted mean density in the group of 5 bins. Large values of $\chi_5^2$ would indicate large deviations from Poisson noise, and if many such bins were found, it would imply significant small spatial-scale variations in the density.
However, contrary to KKF, we find that in the region within $R<20\arcmin$ the observed distribution of $\chi_5^2$ does not differ significantly from the expected distribution; indeed a Kolmogorov-Smirnoff test indicates that the probability that the observed distribution is drawn from the expected $\chi^2$ distribution is 2\%. The hypothesis that the stellar profile of Draco is smooth on small-scales therefore
cannot be confidently rejected. Since any defects in the survey that were not accounted for in our processing will have the effect of driving down this probability, one should
interpret this test as evidence that the stellar profile is smooth.

\subsection{The central profile of Draco}

We include the HST survey in this discussion of the spatial properties
of the Draco dSph, as the resulting deep photometry complements the 
wide-field MegaCam data, giving much better statistics in the centre 
of the galaxy. The CMD derived from the point-sources in the much deeper WFPC2 data is
shown in Fig.~12. (We define point sources to be those objects in Table~5
that have $\chi < 1.5$ and sharpness between $0$ and $0.2$ in both the V and i-bands). The main
sequence turnoff and part of the RGB and HB are clearly visible. 
In this central region of the dwarf galaxy, the number of contaminants to
the point-source catalogue is very small, but we choose to implement a simple 
CMD selection box (as shown in Fig.~12) around the stellar populations of Draco 
to avoid the
obvious contamination. The selection inside the polygon contains a
total of 1122 stars.

\begin{figure}
\begin{center}
\includegraphics[angle=270,width=\hsize]{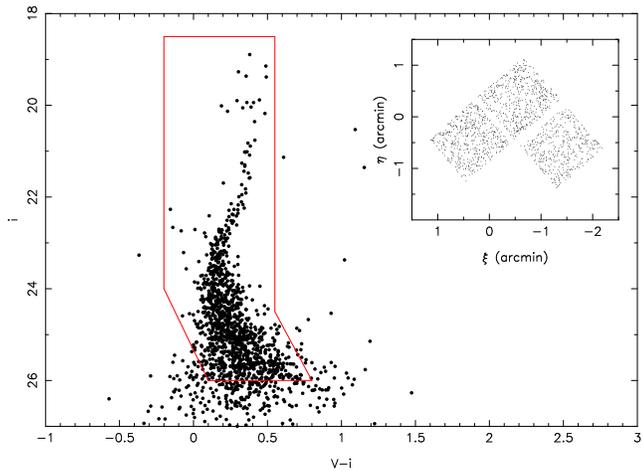}
\caption{The colour-magnitude distribution of stars in the WFPC2 field.
A CMD selection polygon is applied to these data to remove
a small number of probable contaminants. The inset shows the spatial location of these stars at the centre of the dwarf galaxy.}
\label{fig12}
\end{center}
\end{figure}

We now trace the surface density profile of this selection of HST stars. 
We divide the sample into 11 bins each containing 102 stars, and
as with the MegaCam dataset, we compute the stellar density
taking into account the gaps between the CCDs.
Fig.~13 compares the
stellar density profiles of Draco obtained with the HST data to the MegaCam
matched-filter selection. In Fig.~14 we show more clearly the region
probed by the HST data; the higher source density results simply from
the greater photometric depth.
The mean star-count level 
in the HST survey is $296\pm 41 \, {\rm arcmin}^{-2}$, however 
the probability that these density
are constant with radius is 0.8\%. Thus the profile derived from the HST starcounts
requires that the stellar density in the inner regions of the galaxy 
continue to increase towards the centre.

\begin{figure}
\begin{center}
\includegraphics[angle=270,width=\hsize]{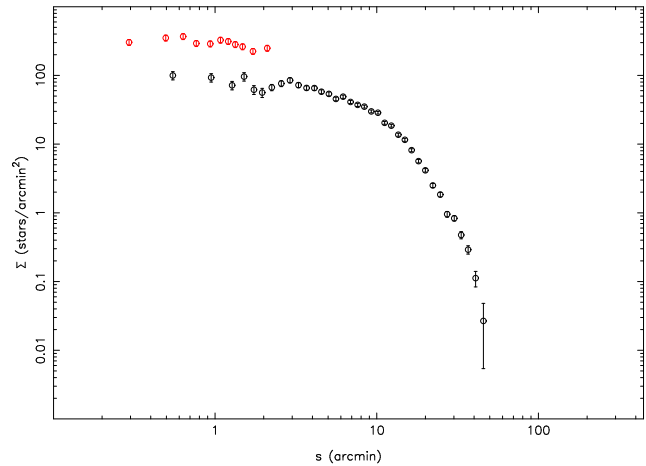}
\caption{The stellar profile of the HST sample (uppermost points), is
compared to the profiled derived from the matched-filter MegaCam
sample, previously displayed in Fig.~9.
The deeper HST data give better statistics in the galaxy centre, allowing us
to probe the density profile down to $s=0\mcnd3$.}
\label{fig13}
\end{center}
\end{figure}

\begin{figure}
\begin{center}
\includegraphics[angle=270,width=\hsize]{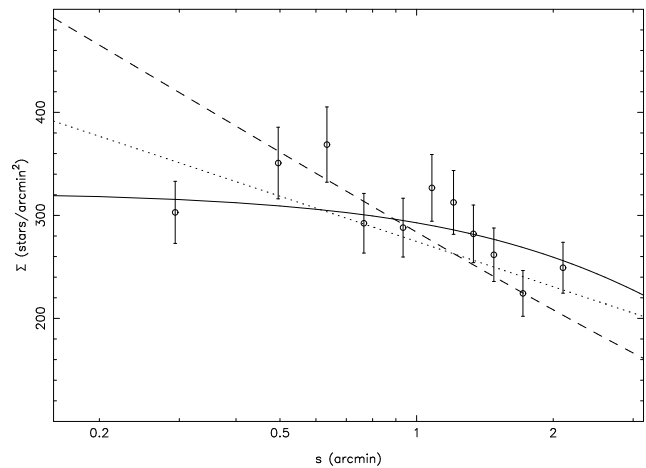}
\caption{The stellar density profile of the central regions of the Draco
dSph, as measured from the HST survey.  Each bin has 102
stars. For comparison, with the dashed line
we display the expected projected
density profile of an NFW model of Virial Mass $10^8\msun$ and concentration
$c=28.6$ (the density normalisation has been chosen to minimise the $\chi^2$
fit to the data). The dotted line shows the profile of an NFW model with $c=4.5$.
The full line shows the exponential model fit previously in the upper panel of
Fig.~10 (but suitably normalised to minimise $\chi^2$).}
\label{fig14}
\end{center}
\end{figure}

In cosmological simulations, dark matter halos follow a "universal"
profile, like the NFW \citep{navarro} or Moore \citep{moore99b}
models. Though these models work well on large scales such as those of
galaxy clusters, the question about the form of the inner regions of
low-mass dark matter halos is not resolved: are their centres cuspy or cored
\citep{deBlok1}? Though the distribution of the luminous matter need not
follow that of the underlying dark halo, it is nevertheless interesting to 
compare the observed distribution of stars to the expected dark matter models.

In Fig.~14 we compare the observed central stellar number density profile to a cusped profile model, 
that we choose to parametrise with a projected NFW model of mass
$10^8\msun$ \citep{lokas01}. We adopt a normalisation
to minimise $\chi^2$, so that the only free parameter is the concentration $c$ 
of the NFW model. The mean concentration of a halo of this mass is $c=28.6$
\citep{maccio}; this model is shown with a dashed line, and has a 
probability of only 0.1\% of conforming to the data. However, 
there is a wide range in the concentration of cosmological halos, with a typical
scatter of 0.4 in $\log(c)$ \citep{maccio}. The dotted line shows a fit for $c=4.5$,
an extreme ($2\sigma$) value for halos of mass $10^8\msun$. The probability
that the observed profile can be drawn from such a model is 8\%.

The marginal discrepancy between the NFW models and the data 
stems solely from the innermost point of the survey.
We have checked this datum carefully, and could find no artificial reason
why it should be substantially lower than the trend. This may reveal a real departure
due to a change in the mass to light ratio of the central regions, though it may also
simply point
to the possibility that the centre of the dwarf galaxy is not precisely
at the location measured by O01. Their centroid resulted from a fit to the data
over the entire galaxy, so it is reasonable that there could be an offset
in the very central $0\mcnd3$. Indeed, this is perfectly consistent
with the quoted uncertainty of $0\mcnd3$ in the O01 centroid position.
The MegaCam survey, due to the large
gaps between CCDs, is not well suited to improve 
on the O01 centroid measurement, and neither is the HST survey presented here
due to its very limited spatial coverage. If we omit the innermost datum, the probability
that the HST profile is drawn from a $c=28.6$ or a $c=4.5$ model is 48\% and 14\%, respectively.

Nevertheless, it is interesting in this context that the simple exponential model that we fit
earlier to the Matched-Filter sample provides a good representation to the data,
including the innermost datum.
This is shown with the full line in Fig.~14, where we have simply scaled the 
previous fit to best match the observed profile. The probability that the data are
drawn from this exponential model is 26\%, while rejecting the innermost point yields a
probability of 28\%.

\subsection{Search for tidal features}

One of the particularities of the Draco dSph based on previous studies of
this system is its undisturbed appearance.  To try to place better
constraints on the absence of tidally-induced features we count the number
of stars at radii between $45\arcmin < R < 120\arcmin$ probing
beyond the King model tidal radius to the end of the survey. Each
annulus is divided into 36 equal parts, and we count the number of stars in
each annulus segment. In Fig.~15 we show the background-subtracted star-counts and
their $1 \sigma$ uncertainties in each annulus segment for the S2 selection.
None of these $10\deg$-wide bins 
exceeds $1\sigma$, so we conclude that no significant overdensities are detected.

To test the sensitivity of this method, we make fake tidal tails by redistributing
a fraction of the stars from the central region of the galaxy along the direction
of the major axis of the galaxy.
For every randomly drawn star from within a $20\arcmin$ radius of the galaxy centre,
we randomly reassign the coordinate along the major axis of the galaxy
so as to have a constant density stream over the length of the surveyed region. 
With a tidal tail representing 3\% of the selection of the central stars, 
we are able to detect the tail at $4.9\sigma$ (summing two bins around $0\deg$) 
and $3.3\sigma$ (from summing two bins around $180\deg$).
This result indicates that a tidal tail
consisting of only 3\% of the central body of Draco
is detectable by this method. 

\begin{figure}
\begin{center}
\includegraphics[angle=270,width=\hsize]{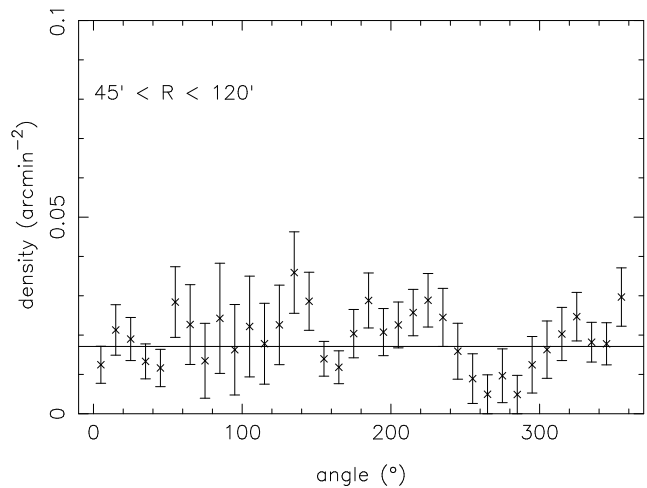}
\caption{Counts in azimuthal bins around the Draco dSph (the
error bars represent 1$\sigma$ uncertainties) for stars beyond $45\arcmin$. 
The data are drawn from
the "S2" selection, which is appropriate for the outer regions of the survey.}
\end{center}
\end{figure}

The absence of any other deformities in the outskirts of the galaxy is further
corroborated by the stellar density map in Fig.~16, which has been constructed
from the matched-filter sample. The uncertainty contours at 
the 1, 3 and $5\sigma$ levels are
also shown. These data clearly show that Draco is a very regular
and symmetric system.

\begin{figure}
\begin{center}
\includegraphics[angle=270,width=\hsize]{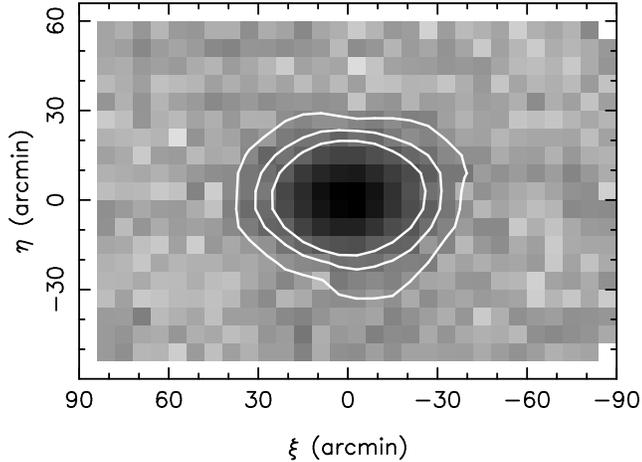}
\caption{Matched-filter map of the MegaCam survey region, shown on a
logarithmic scale and over the same spatial region as in Fig.~6. The 
lines show significance contours of the signal to noise ratio at 1, 3 and 
$5\sigma$.}
\end{center}
\end{figure}

\section{Discussion}

\subsection{The inner density profile}

The HST survey presented above shows clearly that the inner stellar density profile
increases inwards until at least $0\mcnd5$. Inside that radius, the profile
may flatten or decline, or we may simply have adopted a slightly incorrect
centroid. The current data are not good enough to discriminate between these possibilities. 
CDM simulations show dark matter profiles with a
power-law slope between -1 and -1.5, as shown by numerous recent studies
\citep{colin,ghigna,hayashi}. However, low surface brightness galaxies
appear to have flat central mass profiles \citep{deBlok1}, in stark contradiction
to the theoretical predictions. These authors
also show that the profiles are better fit with a model with a power-law
index of $-0.2\pm0.2$.  

A  flat central density profile has also been deduced for the Ursa
Minor dSph, from an analysis of the survivability of a morphological
substructure \citep{kleyna03}. They found that the stars of this substructure are dynamically
colder than the stellar population of the dwarf galaxy, concluding that the
substructure was a stellar cluster accreted by Ursa Minor, and 
whose stars now orbit unbound inside the halo of
the dwarf galaxy. They demonstrated that this
type of scenario is only possible if the dark matter profile around Ursa
Minor has a flat core, as otherwise the stars rapidly lose their coherence.

Some numerical simulations \citep{ricotti} can produce dark matter halos
with flat cores in agreement with those observations. In their cosmological
simulations, \citet{ricotti} created this type of halo at $z\ge 10$, by
taking into account the effect of radiative pressure. They are able to
produce galactic systems similar to the Local Group. To make this, they
selected at $z=10$ galaxies with cored dark matter halos. Then they evolved
without any peroid of fusion. The main result is the core of each dwarf
galaxy remains flat and stable until $z=0$. The modelled velocity dispersion
profile was found to have the same shape as those found observationally by
\citet{kleyna02}.  Another method to obtain dark matter halos with flat core
was proposed by \citet{read}. They discovered it is possible to flatten the
core of a potential by two short periods of mass-loss, separated by a phase
of reaccretion of gas. Recently, \citet{mashchenko} showed that the
CDM cusp may be removed for halos in the $10^7$ to $10^{10}\msun$ range 
by including stellar feedback into the simulations.

Clearly our starcounts are not directly measuring the dark
matter profile; the stellar profile will have resulted from a complex
interplay between baryons and the dark matter. With complementary kinematic measurements we could hope to use the Jeans equations to disentangle the two.
However, though samples of several hundred stars are now being obtained in 
this galaxy \cite{wilkinson04}, the HST photometry we have analysed above probes
the interesting very inner parts of the system, where kinematics are as yet 
unavailable in sufficient numbers to allow a measure of the velocity
dispersion, let alone its gradient. (The paucity of kinematic data is due to the 
low surface number density of bright giant stars needed for 
spectroscopic observations).
Nevertheless,
a clear flattening of the starcounts in the central regions of the galaxy
away from an NFW model would have lent support to the presence of
a core in the Draco dSph. Similarly, if the stars were found
to follow an NFW profile into the very central regions 
it would call into question the possibility that the
dark matter is cored in this dwarf galaxy.
Unfortunately, the uncertainty
in the position of the centroid of the galaxy does not allow us to draw a clear
conclusion on this matter. If we do not include the innermost datum, the 
HST data follow an NFW model closely. Including the innermost datum
forces us to accept a model with a flat inner profile like the exponential model
shown in Fig.~14. Further observational work is required to
measure accurately the position of the centre of the dwarf galaxy, and to improve 
on the statistics of the starcounts in the central few arcmins.

\subsection{Absence of tidal effects}

We have searched for tidal disturbances  in the outer parts
of the  Draco dwarf spheroidal galaxy, using a stellar selection
designed to optimise the signal to noise ratio in regions of low contrast.
However, no evidence was found for the presence of an extra-tidal population,
and we limit the fraction of such a component to less than 3\% within 
the spatial extent of the MegaCam survey. 
This result, in conjunction with the finding that the surface brightness profile
is consistent with that of an unperturbed dynamical system, suggests that Draco 
has led a quiet existence. This finding also lends
support to models in which the stars are shielded by a dark  matter mini-halo
from the tidal forces of the Milky Way.

A similar conclusion about the non-detection of tidal streams in Draco has
been made by \citet{klessen1}, who studied the properties of blue horizontal
branch stars in Draco with the SDSS dataset. They choose this stellar
population because it is distributed all over Draco and the contamination by
other astronomical objetcs is minimal.  Their surface brightness profile of
this stellar population does not possess any break, in agreement with our
much deeper MegaCam results presented here. They completed this study with a
numerical simulation of the dwarf galaxy, without a dark matter component and
affected by the tidal effects of the Milky Way. By comparing the
morphology of the simulation to the observations of the horizontal branch of Draco, they
concluded that such a purely baryonic model is not compatible with reality. Their numerical
simulations show a wide horizontal branch that is not observed for
Draco. Thus \citet{klessen1} conclude that Draco is not a renmant of a
tidally disrupted galaxy, but a dwarf galaxy with a substantial dark matter
halo.

The fact that we find no tidal stream is
not altogether unexpected.  \citet{mayer} conducted cosmological simulations
to attempt to model the behaviour of dwarf galaxies with tidal debris, such
as the Carina dSph. They found that to reproduce this type of galaxy, the
dark matter halo must have a flat core or else it can possess a steep inner
profile but with very low concentration ($<7$). If the concentration is
higher, the orbits must be highly tangential to reproduce the
observations. In the case of dwarf galaxies with high mass-to-light ratio,
like Draco or Ursa Minor, they showed that it is difficult to create large
tidal streams. Only small tidal features could be formed, but at low
contrast.

\section{Conclusions}

We have presented a study of the structure of the Draco dwarf spheroidal galaxy, using
photometric data obtained with the MegaCam camera on the CFHT telescope
and the WFPC2 camera aboard the Hubble Space Telescope.
These two datasets give us a new detailed view of this small galaxy. 

Our deep photometry allows us to probe further than previous studies into the heart and the outskirts of the dwarf galaxy. We find that the central stellar density 
increases inwards from $\sim 2'$ to $0\mcnd5$. We find reasonable agreement if we choose to model
the central HST stellar profile with a cusped model, parametrised as 
a projected NFW model of mass
$10^8\msun$. Inside $0\mcnd5$ the profile appears to flatten out, or drop,
though this may be an artefact of adopting an incorrect value for the position
of the centre of the galaxy. However, we find that an exponential model gives an
equally good description of the data.

The derived stellar profile is very smooth on large scales 
and does not appear to have a 
break, contrary to some previous claims. We also do not find evidence to
support a claim that the stellar distribution is clumpy on small scales.
We fit the stellar profile with various simple models, and improve slightly 
the model parameters compared to earlier studies. However, neither
a King model, an exponential or a Plummer model give satisfactory fits to the observed profile over the entirety of the galaxy. 
The data depart slightly from each of these idealised models, and
we are forced to reject them all at high confidence. However, this 
discrepancy is more 
a testament to the outstanding quality of the data. A more refined model
will have to be developed to properly model these data; this is however beyond the scope of the present paper.

We also search for possible tidal substructures in the outskirts of the dwarf
galaxy. No evidence for the presence of a tidal stream is found. Monte-Carlo tests
show that if any such structure is present it cannot represent more than 3\% of the 
main body of the system. Thus we find that Draco is an undisturbed system,
most likely in dynamical equilibrium.
The Draco dSph therefore continues to be an excellent target for studies of the dark matter on small scales.

\section*{Acknowledgments}

We would like to thank the staff at the CFHT telescope for observing
and processing the MegaCam data, and for their kind help at many 
stages during the progress of this work.

\end{document}